\documentclass[twocolumn,twoside,preprintnumbers,amsmath,amssymb,showkeys]{revtex4}
\usepackage{epsfig}
\usepackage{graphicx}

\usepackage{fancyhdr}

\usepackage{pslatex}

\begin{document}

\title{Screening in the QCD plasma: effects of the gluons and of the confinement}

\author{Giorgio Calucci$^1$, and Enrico Cattaruzza$^2$}

\affiliation{$^1$ Dipartimento di Fisica teorica, Universit\`a di Trieste ed
INFN Sezione di Trieste, Italy \\
         $^2$Dipartimento di Fisica, Universit\`a di Trieste ed
INFN Sezione di Trieste, Italy  }


\begin{abstract} The effects of a thermalized gas of gluons in a $q\bar
q$-plasma is investigated. Then the interplay between Debye screening and confinement is analyzed in a simplified model. While the one-gluon exchange gives results very similar, but not equal, to the electric case, the phenomenological introduction of confinement leads to very different results.

\keywords{Chromodynamical plasma, Confinement}

\end{abstract}
\maketitle

\thispagestyle{fancy}

\setcounter{page}{1}

\section{FORMULATION OF THE PROBLEM}

The existence of a {\it H\"uckel-Debye} color screening is often proposed as 
a signal of the presence of a QCD plasma. It is a property of the electric plasma which is frequently transported to the QCD plasma.\\
In the electric case, with interaction:
 $$ { U=\sum_{i<j}u(|{\bf r}_i-{\bf r}_j|) \quad 
 u(r_{i,j})=\alpha z_i z_j/r\;, }$$
 from the many-body canonical distribution the two, three,$\dots$ body correlation functions are obtained and then 
 the equation for two body correlation is found
{ \begin{align*}
 &\frac {\partial C({\bf r}_1,{\bf r}_2)} {\partial r_{1,v}}=
  -{\beta}\bigg(
    \frac {\partial u({\bf r}_1,{\bf r}_2)} {\partial r_{1,v}}+\\
   \frac {1} {V}& \sum_{l\neq 1,2}\int d^3 r_l\Big[
   \frac{\partial u({\bf r}_1,{\bf r}_l)} {\partial r_{1,v}}
   C({\bf r}_l,{\bf r}_2) \Big]\bigg) 
   \\ &v=x,y,z\;.
   \end{align*} }
   For a macroscopically neutral plasma the result is:
 $$ {C(r) \propto \frac {1} {r}\exp[-ar]\qquad a=\sqrt {\beta\alpha n}\qquad
    n=\frac {N} {V}. }$$
 In QCD the analogous equation for two body correlation is:
   { \begin{align*}
   &\frac {\partial C_{\beta}(r_1,r_2)} {\partial r_{1,v}}= 
   -\int^{\beta}_o d\tau \bigg(
    \frac {\partial u(r_1,r_2)} {\partial r_{1,v}}+\cr
   &\frac {1}{3V}\sum_{l\neq 1,2}\int d^3 r_l\Big[
   \frac {\partial u(r_1,r_l)}{\partial r_{1,v}}
   C_{\beta-\tau}(r_l,r_2)+
   C_{\tau}(r_l,r_2) \frac{\partial u(r_1,r_l)}{\partial r_{1,v}}
   \Big]\bigg)
   \end{align*} 
   \sl (matrix product understood)}.\\
 The color degrees of freedom has two related effects, one is that the correlation functions acquire a matricial shape, the second is
   the integration in $\tau$. It is due to the non commutativity and comes 
 from the representation: {
 $${d\over{dt}}e^A=\int^1_0 e^{xA}{{dA}\over{dt}}e^{(1-x)A} dx\,,$$}verified by series expansion.\\
It is convenient to use {\sl Fourier} and {\sl Laplace} transforms,
 so we get a set of linear equation for the different color components of 
  $\check C$, the {\sl Fourier} and  {\sl Laplace} transform of {$C$}:
\begin{equation*}{ -k^2 \check C(s;{\bf k})={{T({\bf k})}\over {s^2}}+
 {1\over {3Vs}}
   \Big[\sum_{l\neq 1,2} T ({\bf k})\check C(s;{\bf k})+
   \check C(s;{\bf k})T ({\bf k}) \Big], }
   \end{equation*}where
${\bf k}$ is conjugated to the space variables while $s$ is conjugated to the auxiliary variable $\tau$.

\section{QUARK, ANTIQUARK AND GLUON POPULATION}

 At high energy particle population cannot be considered given, but it is the result of the dynamics; we may start from a termal population and then investigate the effects of the interaction. The baryon number $(\times 3)$, when $\epsilon >>m$, is given by:
\begin{equation*}
b=\rho -\bar \rho=\frac{g_f}{2\,\pi^2}\,\int_0^{\infty}d\epsilon\,\epsilon^2
\left[\frac{1}{e^{\beta\,(\epsilon-\mu)}+1}-
\frac{1}{e^{\beta\,(\epsilon+\mu)}+1}\right].
\end{equation*} 
This value is considered as an initial conditiony, in an ion-ion collision it is determined by the baryon number of the colliding particles and also from the total impact parameter, since in a noncentral collision not all the incoming nucleons contribute to the plasma. 
 In particular the density of antiquark is 
\begin{equation*}
\bar \rho=\frac{g_f}{2\,\pi^2}\,
\int_0^{\infty}\frac{d\epsilon\,\epsilon^2}
{1+e^{\beta\,(\epsilon+\mu)}}
=-\frac{g_f}{2\,\pi^2}\,\frac{2}{\beta^3}\,
\mathcal{L}_3\left(-e^{-\beta\,\mu}\right),
\end{equation*} 
where  $\mathcal{L}_3(z)=\sum_{n=1}^{\infty}{z^n}/{n^3}\;.$ \\
 It is possible to express $\mu$ in term of $\beta$ and $b$. By using the
 actual expression of $\rho$ and $\bar\rho$ and the known properties of
 $\mathcal{L}_3(z)$ we get:
\begin{equation*} {
b=\frac{g_f}{6\,\pi^2}\,\mu^3+\frac{g_f}{6\,\beta^2}\,\mu },
\end{equation*}
from which the expression of $\mu$ can be extracted.\\
For gluons, as starting point, we have a black-body density. In the high-energy
limit:$\quad g/n \to 8/9$, where
  $g$ is the gluon density, $n=\rho +\bar \rho$ is the quark density $+$ antiquark density.\\
  At lower temperatures (smaller energy per nucleon) quark-mass corrections arise, on the other hand if the total energy becomes extremely high further flavors, besides $u$ and $d$ could play a role.

          \section{ DYNAMICS}
	  
	 Three types of pair-interaction must be considered:\par
	 \begin{enumerate}
	\item quark-quark, quark-antiquark, antiquark-antiquark interacting
	through gluon exchange;
	\item quark-gluon, antiquark-gluon interacting through gluon
	 exchange;
	 \item gluon-gluon interacting through gluon exchange.	 
	 \end{enumerate}
	Since the main interest is in $qq-$ and $\bar qq-$correlations
	only the terms 1 and 2 will be considered.\\
	Looking in detail to the color structure we see that the interaction terms $T({\bf k})$ may have in the color space the following forms:
	\begin{itemize}
	\item[-]interaction of $qq$ or $\bar q\bar q$
	 $$I_{a,c}^{b,d}=
  	\frac{1}{2}\left[\delta_a^d\,\delta_c^b-
	  \frac{1}{3}\,\delta_a^b\,\delta_c^b\right],$$
	  while    the interaction of $q\bar q$ is given by $-I$;
	\item[-]    interaction of $qg$ 
  $$J_{a,cf}^{b,dg}=\frac{1}{2\,i}
  \left[\delta_a^d\,\delta_c^b\,\delta_f^g-
  \,\delta_a^g\,\delta_c^d\,\delta_f^b\right],$$
while  the interaction of
   $\bar qg$ is given by $-J$.
\end{itemize}
The gluons are in spinorial representation: traceless $G^a_b$.
  The contact term not included, they do not contribute to the {\it relatively}    long distance behavior.\par
The correlation functions $\check C(s;{\bf k})$ may belong to different color representation,  the corresponding color projectors are:
\begin{itemize}
\item[-] for the $qq$, $q\bar q$, $\bar q\bar q$ systems
{\small
 \begin{align*}
{}^3\Pi_{a,c}^{b,d}&=\frac{1}{2}\left[\delta_a^b\,\delta_c^d-\,\delta_a^d\,\delta_c^b\right]&
{}^6\Pi_{a,c}^{b,d}&=\frac{1}{2}\left[\delta_a^b\,\delta_c^d+\,\delta_a^d\,\delta_c^b\right]\nonumber\\
{}^1\Pi_{a,c}^{b,d}&=\frac{1}{3}\,\delta_a^d\,\delta_a^d&
{}^8\Pi_{a,c}^{b,d}&=\left[\delta_a^b\,\delta_c^d-\frac{1}{3}\,\delta_a^d\,\delta_c^b\right];&
\end{align*}}
\item[-] for the $qg$, $\bar q g$ systems
{\small
 \begin{align*} 
{}^3P_{a,cf}^{b,dg}& =\frac{3}{8}\,\delta_{a}^{g}\,\delta_{c}^{d}\,\delta_{f}^{b}-
\frac{1}{8}\,\left[\delta_{a}^{d}\,\delta_{c}^{g}\,\delta_{f}^{b}+\delta_{a}^{g}\,\delta_{c}^{b}\,
\delta_{f}^{d}\right]+\frac{1}{24}\,\delta_{a}^{b}\,\delta_{c}^{g}\,\delta_{f}^{d}\nonumber\\
{}^6P_{a,cf}^{b,dg}& = \frac{1}{2}\,\left[\delta_{a}^{b}\,\delta_{c}^{d}-\delta_{a}^{d}\,\delta_{c}^{b}\right]\,\delta_{f}^{g}-\\
&\frac{1}{4}\,\left[\delta_{a}^{g}\,\delta_{c}^{d}\,\delta_{f}^{b}
+\delta_{a}^{b}\,\delta_{c}^{g}\,\delta_{f}^{d}
-\delta_{a}^{g}\,\delta_{c}^{b}\,\delta_{f}^{d}
-\delta_{a}^{d}\,\delta_{c}^{g}\,\delta_{f}^{b}\right]\nonumber\\
{}^{15}P_{a,cf}^{b,dg}& = \frac{1}{2}\,\left[\delta_{a}^{b}\,\delta_{c}^{d}+\delta_{a}^{d}\,\delta_{c}^{b}\right]\,\delta_{f}^{g}-\\
&\frac{1}{8}\,\left[\delta_{a}^{g}\,\delta_{c}^{d}\,\delta_{f}^{b}
+\delta_{a}^{b}\,\delta_{c}^{g}\,\delta_{f}^{d}
+\delta_{a}^{g}\,\delta_{c}^{b}\,\delta_{f}^{d}
+\delta_{a}^{d}\,\delta_{c}^{g}\,\delta_{f}^{b}\right].
\end{align*}}
\end{itemize}

 The previously given equation for $\check C$ can be projected in the different
 color representations. A system of linear algebraic equations it therefore obtained which can be explicitly solved, but
 the inversion of the {\it Fourier} and {\it Laplace} transform is complicated.
 \par
 For $q\bar q$ correlation $G(r)$ numerical results are found: the damping of the two body correlation is not exponential, but it is not very different from it.
 (This can be seen plotting  $\ln rG(r)$). Moreover at larger distance there is an oscillating behaviour, but only in that region where the damping has already strongly suppressed the correlation.
 \par
 By comparing the result obtained when only real $q$ and $\bar q$ exist (the gluons only mediate the interaction) it appears that
 the presence of real gluons has little qualitative effect. This gives some confidence that the omission of the gluon-gluon interaction term should not have relevant effects; in any case it is evident that its insertion will produce only a heavier task of calculation, not new conceptual problems.\par
 In the case of pure $q \bar q$ population an approximate analytical inversion of the {\it Fourier} and {\it Laplace} transform is possible and it confirms the non exponential decay and also the existence of damped oscillations.
 We can also define a total correlation, by integrating over space the two-body correlation, with the result:
 $$\int d^3r G(r)=\frac {2}{3n}\Big[1+\frac {6g}{n+6g}\Big].$$ 
  The dependence on the temperature is hidden in $g$ and $n$.
 All the results for $qq$ are very similar to the result for $q\bar q$. (Obviously in an ion-ion collision the incoming densities of quark and of antiquark are different).
 
 \section {Confinement}
\subsection {Procedures}

 The dynamical input has been till now perturbative since the potentials were
 generated by one-gluon exchange.
 There is the confinement which has not yet been deduced from gluon exchange.
 But we know from phenomenology that something like the confinement must exist,
 and more precisely we need:\par
 An effective potential in singlet state $q\bar q$ in order to give rise to mesons, and
 from the $\psi$-spectroscopy we have also an indication of its shape:
 $$V_b^a=\delta_b^a\,r/\ell^2$$
 (plus a Coulomb-like sigularity, which is the perturbative term).\par
An effective potential in singlet state $qqq$ in order to give rise to baryons:  $$v^{abc}_{lmn}(r_1,r_2,r_3)=\epsilon^{abc}\epsilon_{lmn}f(r_1-r_2,r_1-r_3).$$
  and here the spatial shape is less well established.\par
  In the case of the three quark state a detailed analysis would require the consideration of the three body correlation function; a very simplified treatment is obtained by using a quark-diquark model. From the point of view of the color an antisymmetric state of two quark is equivalent to a state of an antiquark. The whole procedure is a sort of exaggeration of the confinement effect, in fact every singlet pair feels a confining potential.\\
 From the technical point of view the procedure is the same as in the previous chapter, it is important to keep in mind that the system is anyhow finite, otherways the rising potention would cause unphysical boundary conditions.
  
  \subsection {Results}
 The inversion of the {\it Fourier} and {\it Laplace} transform is studied numerically here also:
 the 2-body correlation shows no damping, but oscillations, moreover they grow with distance. So the result is apparently meaningless. Using the boundary condition that the system is finite it is however 
 possible to define the total correlation (integrated over distance). If there is only confining term, $i.e.$ without the perturbative term, the total correlation is zero, the oscillations compensate spatially.\par
 One could ask which is the origin of oscillation:\par
It is not the color structure: the oscillations are present also in a colorless system with confining potential, where an analytic treatment is possible, so that we are sure that they are not an artifact of the numerical approximations.\par
 It is not the particular power law: they are found also in a colorless system with growing but not linear potential. The general feature of growing potential, not the particular law matters.\par
 Another reason is the existence of a $q\bar q$-potential totally different from the $qq$-potential.\par
 It must be noticed that in this case the behavior of  $q\bar q$ and $qq$ correlation is different.
 
\section{ CONCLUSIONS}

Two aspects of mutual shielding have been examined, the first one amounts to
the inclusion of thermal gluons in the effect of mutual screening in a $q\bar
q-$plasma; it refines a previous analysis given in term of a pure 
quark-antiquark population and it confirms the results. In particular the correlation length 
and the shape of the damping is still the same for $qq$ and $q\bar q$, so that 
the shielding effect would be the same in the meson and in the baryon production, even the oscillations in the tail of the correlation function are equal.\par
The second instance begins with a phenomenological introduction of confining potentials. While the $q\bar q-$interaction is better known, the $qq-$forces present the problem that a real confinement is produced only when three quark, in suitable color state, are put together; this difficulty has been circumvented by using a quark-diquark model. This model is consistent with the interpretation of the confinement as due to a color string and moreover it links the strength of this interaction to the $q\bar q-$interaction. The result has been much more complicated than in the first case. In fact no punctual shielding has been found, what may rise doubts on the whole treament, luckily a spatial integration of the shielding and antishielding effects yields a finite result, so that the final picture we obtain in this way is much more simplified. One could think that this unusual behavior is an oversimplified description of a breaking of the plasma system into smaller droplets, but we lack a detailed description of this possible fact.
\par
The shielding effect depends on the temperature, this dependence
has two origins: one is the direct effect that is present in every
plasma-like system, the other, typical of a relativistic system, comes from the thermal production
of particles, both gluons and $q\bar q$ pairs. Initially there are more quarks than
antiquarks; however, looking at the evolution of the chemical potential, we see 
that there is an equivalent temperature, not extremely high, at which the
effects of the initial condition become very small.\\
Some words of comparison with other ways of
dealing with the same phenomenon. The treatment given in term of strong coupling
on the lattice is difficult to compare with, it is very different
since the beginning; more similar are the treaments in terms of thermal Green
function. If in that case one extracts the contribution of
the pole of the propagator in momentum space, a precise Yukawa-like decay 
of the correlation function is certainly produced. 
The more unusual result, $i.e$ the presence of an antishielding, was already foreseen, but in the different situation of an anisotropic plasma.\\  
We have presented here a particular and definite model in which some of the dynamical features, the static interactions, 
are put into evidence and worked out until they yield a quantitative result. 
 
\noindent{\bf Acknowledgements:} 
Presented at ISMD2006, Paraty (R.J.) Brazil, September 2006.
\par
 In order to save space only the papers used in producing this presentation
 have been quoted, in them the relevant references can be found.


\end{document}